# Quantum paramagnet near spin-state transition

*Keisuke Tomiyasu,*[1]* *Naoko Ito,*[2] *Ryuji Okazaki,*[2] *Yuki Takahashi,*[1] *Mitsugi Onodera,*[1] *Kazuaki Iwasa,*[1,3] *Tsutomu Nojima,*[4] *Takuya Aoyama,*[1] *Kenya Ohgushi,*[1] *Yoshihisa Ishikawa,*[5] *Takashi Kamiyama,*[5,6] *Seiko Ohira-Kawamura,*[7] *Maiko Kofu,*[7] *and Sumio Ishihara*[1]

[1]Department of Physics, Tohoku University, Aoba, Sendai 980-8578, Japan
[2]Department of Physics, Faculty of Science and Technology, Tokyo University of Science, Noda, 278-8510, Japan
[3]Frontier Research Center for Applied Atomic Sciences, Ibaraki University, Tokai, Ibaraki 319-1106, Japan
[4]Institute for Materials Research, Tohoku University, Aoba, Sendai 980-8577, Japan
[5]Institute of Materials Structure Science, KEK, Tokai, Ibaraki 319-1106, Japan
[6]Sokendai (The Graduate University for Advanced Studies), Tokai, Ibaraki 319-1106, Japan
[7]J-PARC Center, KEK, Tokai, Ibaraki 319-1106, Japan
*E-mail: tomiyasu@tohoku.ac.jp

Keywords: magneto-lattice-expansion, excitonic insulator, bi-exciton, spin-state transition

*ABSTRACT*
Spin-state transition, also known as spin crossover, plays a key role in diverse systems, including minerals and biological materials. In theory, the boundary range between the low- and high-spin states is expected to enrich the transition and give rise to unusual physical states. However, no compound that realizes a nearly degenerate critical range as the ground state without requiring special external conditions has yet been experimentally identified. This study reports that, by comprehensive measurements of macroscopic physical properties, X-ray diffractometry, and neutron spectroscopy, the Sc substitution in $LaCoO_3$ destabilizes its nonmagnetic low-spin state and generates an anomalous paramagnetic state accompanied by the enhancement of transport gap and magneto-lattice-expansion as well as the contraction of Co–O distance with the increase of electron site-transfer. These phenomena are not well described by the mixture of conventional low- and high-spin states, but by their quantum superposition occurring on the verge of a spin-state transition. The present study enables us to significantly accelerate the design of new advanced materials without requiring special equipment based on the concept of quantum spin-state criticality.

Spin-state transitions appear in a wide range of sciences.[1] For example, they naturally occur under the high pressure of the Earth's inner mantle, and are related to seismic-wave responses.[2,3] Spin-crossover complexes exhibit magneto-optical switching functionality.[4] Heme-protein activities involve spin-state transitions.[5] The spin-state transition denotes a change between two well-defined spin states, whose boundary ranges provide an immense scope for fascinating phenomena. Conventional low-spin (LS) and high-spin (HS) states (**Figure 1**) are known to exist in compounds of octahedrally coordinated $3d^6$ transition metal ions ($Co^{3+}$ and $Fe^{2+}$). The crystal-field energy stabilizes the LS state, whereas the intra-atomic Hund's energy stabilizes the HS state. When the energy gap between the two spin states ($E_g$) is narrow, an unconventional $Co^{3+}$ spin state described by the quantum superposition of the different spin states is possible. An excitonic-insulating (EI) state has also recently been theoretically predicted to emerge when the quantum spin state is effectively itinerant.[6,7]

A platform compound, which is a promising candidate for the realization of the LS–HS critical range as the ground state, is the perovskite-type oxide, $LaCoO_3$, which is known as the archetypal system that exhibits thermal spin-state transitions. Its $Co^{3+}$ ions are in a ground nonmagnetic LS state at the minimum temperature, and they begin to become thermally activated to a magnetic HS state above a temperature of approximately 30 K,[8–11] indicating a sensitive spin-state variability near the spin-state critical range. Both the application of ultra-high magnetic fields stronger than 100 T and the fabrication of thin films can result in spin-state orders (different spin states are spatially ordered) accompanied with ferromagnetism, which provides direct evidence of the close proximity of $LaCoO_3$ to spin-state instability, and could possibly be related to some EI nature.[12,13] However, few compounds have yet been recognized as a spin-state criticality in the ground state without requiring special external conditions, which hampers the design of advanced materials with respect to the spin-state criticality and the experimental clarification of its potential functions.

$Pr_{0.5}Ca_{0.5}CoO_3$ and related cobaltites exhibit a metal–insulator transition accompanied by the changes of Pr–Co valences and Co spin states, which is claimed to be the EI transition without requiring special external conditions.[6] However, the Co valences are in fairly hole-doped states (typically 3.2+).[27] In addition, the Co-$3d$ electrons are considerably coupled with the Pr-$4f$ electrons.[28] These characteristics may experimentally obfuscate the EI effect in the transition.

In this study, we shed light on the novel substitution system, $LaCo_{1-y}Sc_yO_3$, as another pathway based on the following strategy: (1) $La^{3+}$ possesses no $4f$ electron and Sc (group 3 element) is normally stable at 3+; hence, the Sc substitution directly affects the Co electronic state with a maintained $Co^{3+}$ ($d^6$) matrix; (2) the Co-site substitution effect is explained by Pauling's electronegativity ($\chi_{EN}$);[24] $\chi_{EN}$(Sc) = 1.36 is much smaller than $\chi_{EN}$(Co) = 1.88; hence, the Sc atoms will repulse the ligand O electrons and increase the Co–O covalence, leading to the Co–O–Co electron transfer;[24] and (3) therefore, the Sc substitution is expected to cause the broadening of the Co-$3d$-orbital bandwidth $W$ and the narrowing of $E_g$.

***RESULTS.—  A. Macroscopic fundamental physical properties.*** **Figure 2**A shows the temperature ($T$) dependence of the magnetic susceptibility ($\chi$). As $y$ increased, the nonmagnetic low-$T$ range of the LS state narrowed until $y$ = 0.04, at which $\chi$ began to significantly increase on cooling. No magnetic anomaly was observed for $T$ values as low as 1.8 K or for $y$ values as high as 0.10, indicating the occurrence of a nonmagnetism-to-paramagnetism transition caused by the Sc substitution. The zero-field-cooling (ZFC) and field-cooling (FC) curves slightly split at $T \sim 60$ K when $y$ = 0.125, indicating that a weak static magnetic correlation grows.

Figure 2B shows the $T$ dependence of the dilatability defined as non-dimensional $\varepsilon = \Delta L/L$ without a magnetic field, where $L$ denotes the linear dimension of the sample and $\Delta L$ represents the variation. No anomaly was detected in any of the samples until $T$ = 1.8 K as well as $\chi(T)$. The $y$ = 0 sample rapidly shrunk toward the lowest-$T$ plateau as $T$ decreased below approximately 80 K, corresponding to a well-stabilized LS range.[14] This plateau range narrowed with the increasing values of $y$ and eventually disappeared at $y$ = 0.04.

Figure 2C shows the $T$ dependence of the electrical resistivity ($\rho$). All of the samples showed an exponential increase in $\rho$ with the decreasing $T$, and, hence, were insulating in the measured $T$ range. $\rho$ did not show a systematic increase during the Sc substitution, indicating that the substituted Sc had a stable electronic configuration of $Sc^{3+}$ ($3d^0$, nonmagnetic) without doping of the $Co^{3+}$ matrix with charge carriers, as expected. Figure 2D shows the $y$ dependence of the transport gap ($E_g^S$ and $E_g^\rho$) obtained from the thermopower ($S$) and $\rho$ data, respectively, for which the Supporting Information summarizes the measured $S$ data, the performed data-fitting procedure, and other detailed discussions. Both $E_g^S$ and $E_g^\rho$ commonly showed an upturn anomaly at approximately $y$ = 0.04, implying a change in the electronic structure toward the insulating side at values above $y$ = 0.04. Note that a similar enhancement of the insulating nature was observed in the pressure-induced EI state.[15]

Figure 2E shows the magnetic field ($H$) dependence of the magnetization ($M$) measured at $T = 1.8$ K. $M$ also significantly increased from $y = 0.04$ as $y$ increased. The $y = 0.04$ sample exhibited a response that was approximately linear, as shown by the bold straight line. The Brillouin-like saturating behavior gradually grew when $y$ further increased above 0.05. The $y = 0.125$ curve exhibited a limited remnant magnetization in concurrence with the ZFC-FC splitting of $\chi(T)$.

Next, we measured the response of the dilatability to $H$ (the magneto-lattice-expansion $\varepsilon(H) = \Delta L(H)/L$) at $T = 1.8$ K, as shown in **Figure 3**A. For the well-stabilized LS state in the $y = 0$ sample, a small $\varepsilon$ value of only $2 \times 10^{-5}$ was observed at $H = 7$ T. However, as $y$ increased, a considerably large $\varepsilon$ value of $1 \times 10^{-4}$ at $H = 7$ T was observed above $y = 0.04$, which was consistent with the enhancement of $\chi$, $M$, and transport gap. Furthermore, $\varepsilon(H)$ was described by an even function of $H$, which is generally obtained in the case that time reversal symmetry is conserved.[16] This result was also consistent with the observation that LaCo$_{1-y}$Sc$_y$O$_3$ was basically paramagnetic.

In addition, we obtained $\varepsilon(H)$ measurements for antiferromagnetic Sr$_2$CoO$_3$Cl at $T = 1.8$ K (with a HS reference),[17] paramagnetic LaCoO$_3$ at $T = 300$ K (with a mixed LS-HS reference), and LaCoO$_3$ at $T = 50$ K (on the verge of exhibiting the HS state),[10,11] as shown in the Supporting Information. All the reference data exhibited maximum values for the magneto-lattice-expansion of only $2 \times 10^{-5}$ at $H = 7$ T; hence, contrary to the LaCoO$_3$, LaCo$_{1-y}$Sc$_y$O$_3$ will not be understood that the HS state appears in the LS matrix.

The $H$-parallel $\varepsilon$ data were considered up to this point. We compared these data to the $H$-perpendicular $\varepsilon$ data (Figure 3A, inset). Both $\varepsilon$ values were positive, and had similar magnitudes, indicating that the lattice expanded in an isotropic manner upon applying an external magnetic field, primarily demonstrating volume magneto-lattice-expansion instead of the Jahn–Teller-like anisotropic deformation. Figure 3B shows $\varepsilon(H)$ measured at the elevated $T$ values for the representative $y = 0.05$ sample, which gradually evolved with the decreasing $T$ values. The inset of Figure 3B shows the $T$ dependence of $\varepsilon(7\mathrm{T})$, which was remarkably large below approximately $T = 30$ K.

Figure 3C shows a contour map presenting a full landscape of the magneto-lattice-expansion $\varepsilon(7\mathrm{T})$ values along with the LS range estimated by $\chi(T)$ and $\varepsilon(T)$. This appearance was highly analogous to the left part of Figure 1. In the low-$T$ range, as $y$ increased above $y = 0.04$, the LS state disappeared, and a newly found unconventional paramagnetic state accompanied by the enhancement of transport gap and magneto-lattice-expansion evolved. Interestingly, the magneto-lattice-expansion, which is typically studied for ferromagnetic compounds,[18] is effective as a probe to characterize a paramagnetic system.

*B. X-ray diffraction.* We measured the lattice parameters at $T = 4$ K using powder X-ray diffraction. Both the experimental data obtained and the data fitting procedure performed were summarized in the Supporting Information. The resulting $y$ dependencies of the lattice volume and the averaged $B$–O distance are shown in **Figure 4**A, where $B$ = Co or Sc. As $y$ increased, the volume and the distance became appreciably smaller than the reference lines provided by Vegard's law. These lines were also directly obtained by measuring the $y = 0$ and $y = 1$ samples at $T = 4$ K. They correspond to a mixture of Co$^{3+(LS)}$ and Sc$^{3+}$ because the $y = 0$ sample completely remains within an LS state in the low-$T$ range. This behavior was in contrast to the nonmagnetic reference LaAl$_{1-y}$Sc$_y$O$_3$ that obeys Vegard's law,[20] in which Al$^{3+}$ possesses almost the same ionic radius as Co$^{3+(LS)}$ ($r[\mathrm{Al}^{3+}] = 0.535$ Å, $r[\mathrm{Co}^{3+(LS)}] = 0.545$ Å[19]). Thus, the Co–O distance was contracted in LaCo$_{1-y}$Sc$_y$O$_3$; hence, it will be also unlikely that the HS state appears in the LS matrix in terms of the ionic radius ($r[\mathrm{Co}^{3+(HS)}] = 0.61$ Å $> r[\mathrm{Co}^{3+(LS)}]$[19]).

Figure 4B shows the $y$ dependencies of the obtained angles $\angle(B$–O–$B)$ and $\angle(\mathrm{Co}$–O–Co$)$. Although the accuracy was somewhat low because of the X-ray insensitivity to oxygen, $\angle(B$–O–$B)$ was certainly larger than the reference line. Furthermore, as $y$ increased, $\angle(\mathrm{Co}$–O–Co$)$ was around and above the upper limit value of 162.8° that the material can maintain the nonmagnetic LS state. First-principle calculations and structural experiments for the thermal change of $LaCoO_3$ characterized the crossing of this limit value as the nonmagnetic-to-magnetic spin-state boundary.[21] This fact also suggests that $LaCo_{1-y}Sc_yO_3$ is located in the spin-state boundary range as the ground state.

*C. Neutron spectroscopy.* We performed neutron spectroscopy experiments. **Figure 5**A shows the data measured for the $y = 0$ parent material ($LaCoO_3$). At the lowest $T = 5$ K, no magnetic response was observed reflecting the non-magnetism. When the magnetic spin states were thermally activated at $T = 50$ and 100 K, gapped magnetic excitations appeared at 0.6 meV. This behavior was consistent with the previous report, which observed the overall temperature dependence of the integrated intensity (Figure 5B).[9]

In contrast, Figure 5C shows the data measured for the $y = 0.05$ composition as the representative of the newly found state. The clear gapped excitations were already observed at the lowest $T = 5$ K. This microscopically also indicates that the $y = 0.05$ composition comes out of the non-magnetism as the ground state. Furthermore, when the temperature increased, the spectrum width was widened, but the intensity was substantially large at $T = 300$ K. The temperature dependence of the integrated intensity extracted from the spectra by using the Gaussian fitting was added in Figure 5B, which is distinctive in comparison with the reference $y = 0$ line. Further analysis and discussion were described in the next section.

*DISCUSSION.—* We searched for a model of the newly discovered state. It is widely accepted that $LaCoO_3$ ($y = 0$) is energetically described by a two-level system: the ground LS state and the excited HS state (effective total angular momentum $J_{eff} = 1$ and $g$ factor $g \sim 3.4$) with an energy gap of $E_g \sim 10$ meV.[9,22] This approximate scheme is theoretically verified to be applicable to both the atomic and collective pictures, in which the energy gap denotes the average in the system.[23] Meanwhile, as mentioned in the Introduction section, the high Co–O covalence is expected in $LaCo_{1-y}Sc_yO_3$ from Pauling's electronegativity ($\chi_{EN}$),[24] which will contract the Co–O distance and increase the Co–O–Co electron transfer. Indeed, experimentally, the averaged $B$–O distance became appreciably smaller than Vegard's line (Figure 4A), and the averaged $\angle(B$–O–$B)$ angle was closer to 180° than the reference line (Figure 4B). Hence, the energy gap $E_g$ will be narrowed, which will give rise to the observed nonmagnetic-to-paramagnetic change.

However, the model consisting of the conventional LS and HS states was insufficient to explain the enhancement of transport gap and magneto-lattice-expansion and the contraction of Co-O distance, as mentioned in the Results section. Furthermore, the $\varepsilon$-$M$ correspondence was not obtained, as explained below. The $\varepsilon(H_0)$ value is described as follows:

$$\varepsilon(H_0) = \Delta r(H_0)/r(0) \approx (0.12 \pm 0.02)\Delta p_{HS}(H_0),$$

where $\Delta r(H_0) = r(H_0) - r(0)$, $r(H) = p_{HS}(H_0)r_{HS} + p_{LS}(H_0)r_{LS}$ (an averaged ionic radius of $Co^{3+}$); $p_{HS}$ denotes the probability of the HS state; $\Delta p_{HS}(H_0) = p_{HS}(H_0) - p_{HS}(0)$; and the final deformation is obtained using Shannon's ionic radii. Please see the Supporting Information for details. Hence, for example, the observed $\varepsilon(7T) = 9 \times 10^{-5}$ for $y = 0.04$ (Figure 3) will correspond to $\Delta p_{HS}(7T) \approx 8 \times 10^{-4}$. However, this will induce magnetization $M_{max} = \mu_{HS} \Delta p_{HS}(7T) = 3 \times 10^{-3}$ $\mu_B$/f.u. at maximum, which is one-order smaller than the measured $M(7T) = 3 \times 10^{-2}$ $\mu_B$/f.u. (Figure 2E). For $\mu_{HS} = gJ_{eff}\mu_B$, where $\mu_B$ denotes the Bohr magneton, we used the HS values of $g = 3.4$ and $J_{eff} = 1$, which were experimentally determined for $LaCoO_3$.[9,10,22]

Thus, we attempted to attain another model a step further. The quantum superposition between LS and HS states is possible when $E_g$ is small. Therefore, the following expanded model will become a natural candidate for describing $LaCo_{1-y}Sc_yO_3$. That is, the ground LS and excited HS states energetically approached each other as $y$ increased, and above $y = 0.04$, transformed to the LS-based HS-mixed and HS-based LS-mixed quantum states, respectively.

The LS-based quantum state is described by the linear combination of:

$$|\text{LS-Q}\rangle = (|\text{LS}\rangle + \alpha|\text{HS}\rangle)/\sqrt{1 + |\alpha|^2} = (1 - |\alpha|^2/2)(|\text{LS}\rangle + \alpha|\text{HS}\rangle) = |\text{LS}\rangle + \alpha|\text{HS}\rangle$$

where $|\alpha| \ll 1$. Under $H$, it will change to $|\text{LS-Q}'\rangle = |\text{LS-Q}\rangle + (\alpha\mu_{\text{HS}}H/E_g)|\text{HS}\rangle$ by Van Vleck's formula[25] even when $H$ and $T$ are low to activate the excited magnetic states. The $H$ term indicates that the HS characteristics become denser with applying $H$, resulting in a larger magnetic moment and a larger ionic radius than the LS state. Thus, the LS-based quantum state can generate the simultaneous enhancement of magnetization and volume expansion under $H$.

The temperature dependence of magneto-lattice-expansion ($\varepsilon$) was well analyzed in this model, as shown by the black solid curve in the inset of Figure 3B. The detailed procedure for the analysis was summarized in the Supporting Information. The resultant $E_g$ value was 2.9 meV, indicating that the Sc substitution significantly narrowed $E_g$ from the $y = 0$ value of 10 meV. Furthermore, the magnitude of $\alpha$ was estimated below. This value is connected to the $H$-linear component in magnetization ($M$) caused by the Van Vleck paramagnetism,[7,25] in which the linear susceptibility at the lowest temperature is described by $\chi_{\text{VV,cal}} = 4|\alpha|^2\mu_{\text{HS}}^2/E_g$ (please see the Supporting Information for details). Meanwhile, experimentally, the $y = 0.04$ magnetization curve was approximately linear in all the present samples (Figure 2E). Therefore, $\chi_{\text{VV,exp}} = 2.8 \times 10^{-3}$ $\mu_B$ f.u.$^{-1}$ T$^{-1}$ at $T = 1.8$ K was estimated by assuming this slope as the Van Vleck component. Accordingly, $|\alpha| = 0.06$ was roughly obtained by using the equality $\chi_{\text{VV,cal}} = \chi_{\text{VV,exp}}$ and the aforementioned $E_g = 2.9$ meV. This supports the claim regarding $|\alpha| \ll 1$. Further analysis and discussion in the quantum model were provided in the Supporting Information.

Having identified the appearance of the LS-based quantum state, we further examined the energy level scheme with the neutron spectroscopy data. The intensity for the energy transfer $\hbar\omega$ from $|\Gamma_n\rangle$ with energy $E_n$ to $|\Gamma_m\rangle$ with energy $E_m$ is described by[9]

$$I(\hbar\omega) = C \exp\left(-\frac{E_n}{k_BT}\right)/Z(T)\, F^2(Q)\, |\langle\Gamma_m|M_\perp|\Gamma_n\rangle|^2\, \delta(E_m - E_n - \hbar\omega),$$

where $C$ denotes the scale constant of intensity; $Z(T)$ is the partition function; $k_B$ is the Boltzmann constant; $F(Q)$ is the magnetic form factor; and $M_\perp$ is the component of magnetic moment perpendicular to the scattering vector $Q$ and the delta function is the energy conservation law, which is experimentally relaxed to Gaussian or Lorentzian. The matrix elements of $|\langle\Gamma_m|M_\perp|\Gamma_n\rangle|^2$ for both cases of $|\Gamma\rangle = |\text{LS}\rangle, |\text{HS}\rangle$ and $|\Gamma\rangle = |\text{LS-Q}\rangle, |\text{HS-Q}\rangle$ were summarized in the Supporting Information. With this equation, the $y = 0$ reference line (Figure 5B) was connected to the $y = 0$ energy scheme (left side, Figure 5D).[9] Meanwhile, considering that the energy scheme will have a distribution depending on the distance from Sc in LaCo$_{1-y}$Sc$_y$O$_3$, the $y = 0.05$ data (Figure 5B) will be described by the expanded form:

$$I_{\text{total}}(\hbar\omega) = \int_{-\infty}^{\infty} dE_g\, P(E_g)\, I(\hbar\omega),$$

where $P(E_g)$ denotes the existence probability of the Co sites with $E_g = E_{\text{HS-Q}} - E_{\text{LS-Q}}$, and both the positive and negative signs are allowed for $E_g$ corresponding to the Co sites that exhibit the ground LS- and HS-based quantum states, respectively. Furthermore, for simplicity, we represent $P$ by the normal distribution with the mean $E_g$ value $\langle E_g \rangle$ and the standard deviation $\Delta E_g$. Thus, we obtained the black solid curve by the least-square fitting (Figure 5B) and the parameters of $\langle E_g \rangle = 2.0$ meV and $\Delta E_g = 2.3$ meV (Figure 5D). The curve was in agreement with the experimental data. The mean value $\langle E_g \rangle = 2.0$ meV was also consistent with the value of 2.9 meV estimated from the magneto-lattice-expansion data, and the deviation $\Delta E_g = 2.3$ meV moderately covered the LS–HS degenerate point.

Finally, we discuss the local and itinerant (EI) nature for the spin state. The neutron scattering experiments observed non-dispersive gapped excitations, suggesting the local nature. In contrast, the enhancement of the transport gap suggests the EI nature (Figure 2D).[15] The X-ray diffraction experiments also structurally suggested the increase of the electron transfer (itinerancy). We consider a following model to resolve these situations. In the EI theories, the intermediate-spin (IS) state generates the spatially mobile exciton as the source of dispersive spin-state excitations.[6,7] By contrast, the HS state is described as an immobile bi-exciton consisting of two types of IS excitons with the different site-hopping directionality and results in the non-dispersive excitations.[6c),30] In this scheme, the HS condensation ranging multiple sites would lead to a novel bi-excitonic insulating (bi-EI) state with dual nature of immobility (localization) and site-coherence (itinerancy). Furthermore, a recent study by neutron scattering and theoretical calculations reported that the local HS excitations in parent $LaCoO_3$ form a very short-ranged (approximately seven-Co-sites) collective unit involving delocalized/itinerant electrons,[29] which would be a seed of the bi-EI state with dual nature in $LaCo_{1-y}Sc_yO_3$. Thus, future studies to clarify the spin states in $LaCo_{1-y}Sc_yO_3$ will be intriguing.

In conclusion, $LaCo_{1-y}Sc_yO_3$ achieves the spin-state critical range as the ground state without requiring special external conditions, such as film fabrications, high pressures, and high magnetic fields. Furthermore, this range is characterized as a novel quantum paramagnet accompanied by the enhancement of the transport gap and volume magneto-lattice-expansion as well as the contraction of Co–O distance, indicating that $LaCo_{1-y}Sc_yO_3$ is a highly promising candidate for the new spin-state bi-EI and/or EI related material. Thus, these results enable us to significantly accelerate the design of advanced materials in the laboratories without requiring special equipment based on the concept of quantum spin-state criticality. The materialization of the EI state has been an issue of interest for more than half a century because it is theoretically based on a zero-charge hole-electron Cooper pair.[26] Therefore, the spin-state instability would also provide a novel pathway for superconductivity, quantum spin-state liquids, and magneto-effects such as the magneto-lattice-expansion presented by this work.

**Acknowledgments**
We would like to thank Prof. N. Kimura of Tohoku University for his technical help. The neutron experiments were performed with the approval of J-PARC (2017A0268). This study was financially supported by the MEXT and JSPS KAKENHI (JP18K03503, JP17H06137, JP15H03692, and JP17H02916) and by the FRIS Program for the creation of interdisciplinary research at Tohoku University.

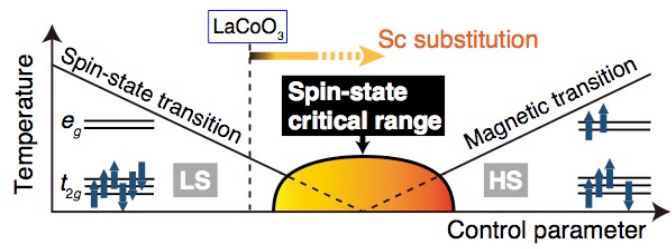

**Figure 1.** Conceptual design of the control of the spin-state transition. The left and right sides show the spin-orbital schemes of LS and HS states, respectively. The six 3$d$ electrons occupy the triply degenerate $t_{2g}$ orbital and the doubly degenerate $e_g$ orbital in different manners. The central spin-state critical range denotes that the novel spin-orbit states, such as the quantum linear combination states of the different spin-states, are theoretically possible on the verge of the LS and HS degenerate point. The Sc substitution into LaCoO$_3$ acts as a control parameter and is expected to push the system toward the critical range.

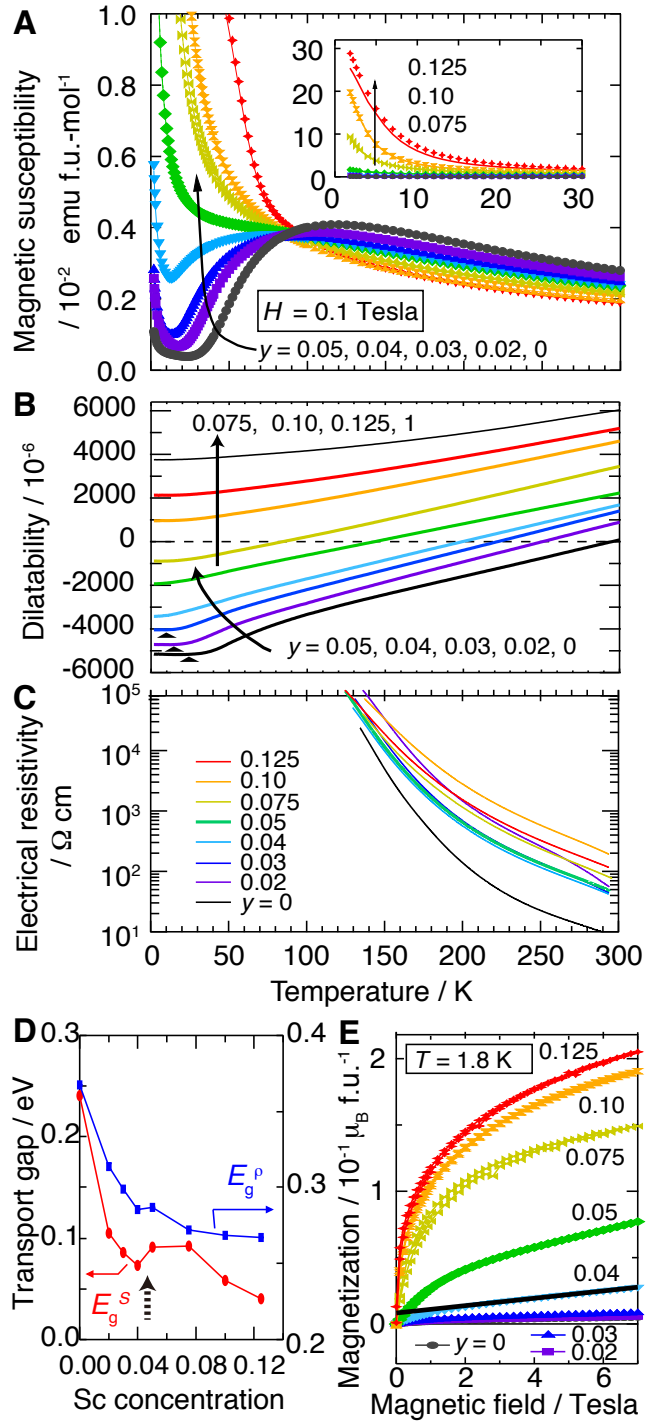

**Figure 2.** Fundamental physical properties. A) Temperature dependence of magnetic susceptibility. The ZFC and FC curves are plotted using curves and symbols, respectively, and trace one another except at $y = 0.125$. The inset panel shows the high-susceptibility range for the low-$T$ range. B) Temperature dependence of dilatability. The three triangles in the bottom left corner denote the kinks corresponding to the completion of the spin-state transition to the LS state. The data are offset vertically for clarity. C) Temperature dependence of electrical resistivity. D) Sc concentration dependence of the transport gap obtained from thermopower ($S$) and electrical resistivity ($\rho$). E) Magnetization process. The black solid straight line shows that the $y = 0.04$ curve is approximately linear.

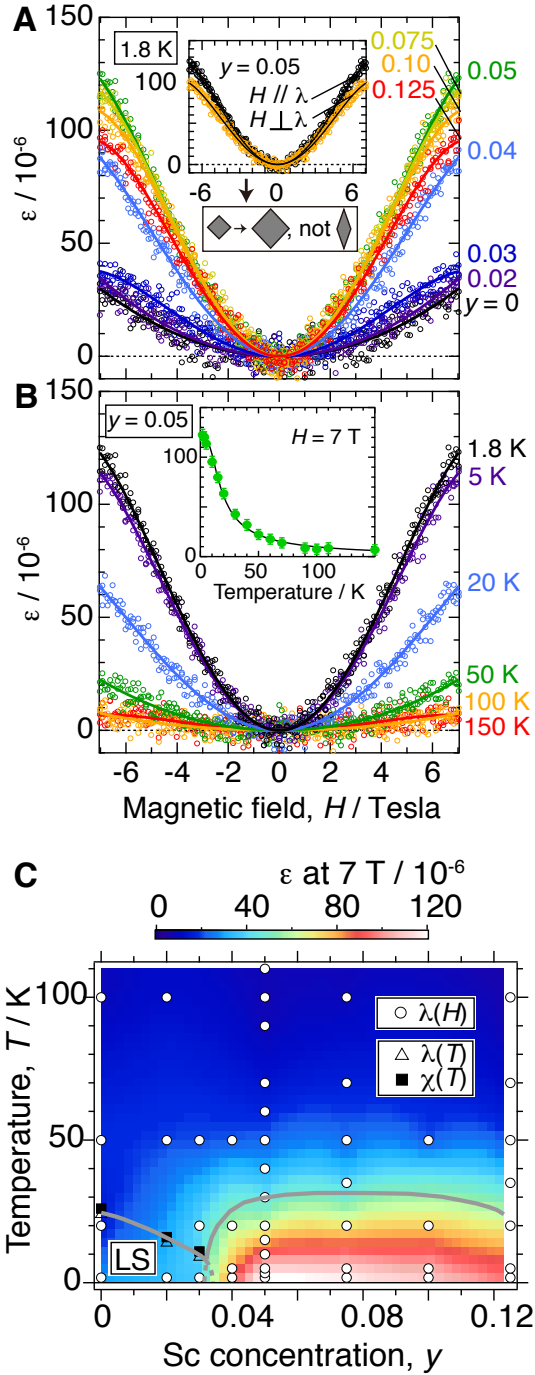

**Figure 3.** Magneto-lattice-expansion. A) Magneto-lattice-expansion, $\varepsilon(H)$, measured at $T = 1.8$ K for all the $y$ samples. The small circles denote the experimental data vertically shifted to zero at $H = 0$. The curves are obtained by even-parity $H^2$ and $H^4$ polynomial fitting. The inset shows a comparison between the longitudinal and transverse magneto-lattice-expansions. The diamonds demonstrate the volume dilation type that occurs. B) Temperature evolution of the magneto-lattice-expansion measured for the $y = 0.05$ sample. The inset shows the temperature dependence of the values observed at 7 T, while the curve demonstrates the fitting result (see the Supporting Information). C) Summarized contour map of the magneto-lattice-expansion observed at 7 T [$\varepsilon(7T)$]. The open circles indicate the measured $y$-$T$ points, while the gray curves are provided as visual guides. The LS range estimated by $\varepsilon(T)$ and $\chi(T)$ is superimposed.

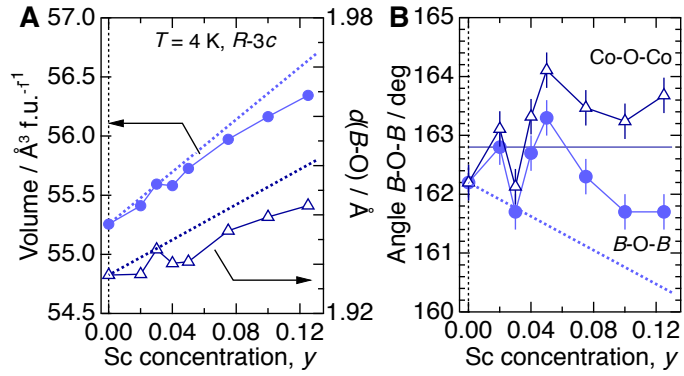

**Figure 4.** Lattice parameters determined by X-ray diffraction at $T = 4$ K. Sc concentration dependencies of A) the lattice volume and the averaged $B$–O distance and B) the angles $B$–O–$B$ and Co–O–Co. The errors are smaller than the symbol sizes when no error bars are shown. The straight dotted lines denote the reference lines directly connecting the $y = 0$ and $y = 1$ data measured at this temperature. In (B), the horizontal straight thin line denotes the upper limit that the LS state can be maintained, 162.8°.[21]

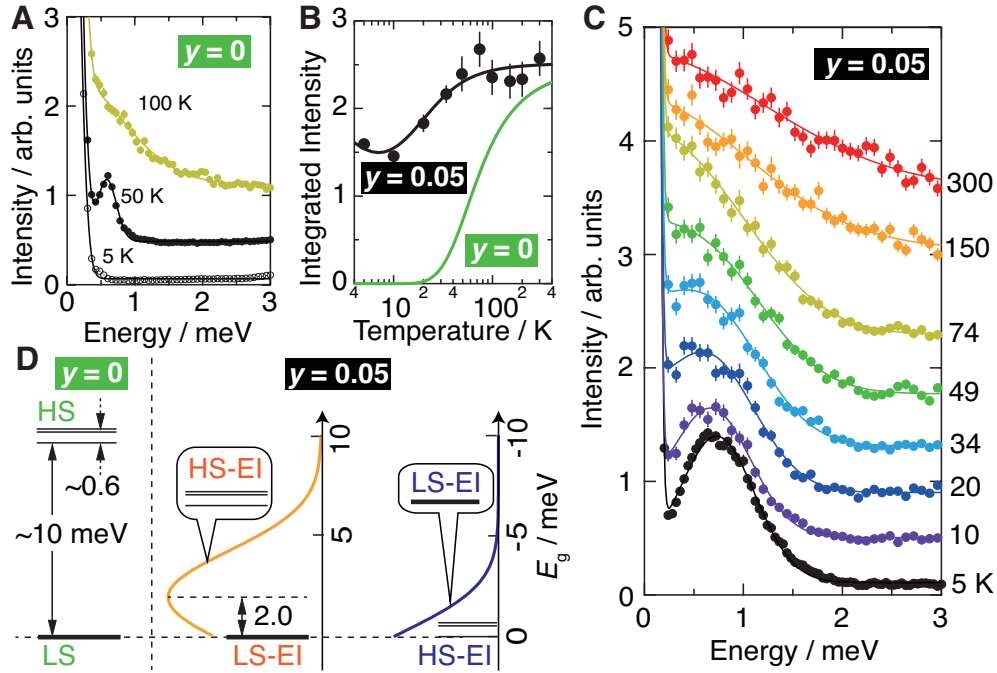

**Figure 5.** Neutron spectroscopy data. A) Energy spectra measured for the $y = 0$ composition. B) Temperature dependence of the integrated intensity of the 0.6 meV excitations. C) Energy spectra measured for the $y = 0.05$ composition. D) Energy level schemes obtained for the $y = 0$ and $y = 0.05$ compositions. The curves show the function of existing probability, $P(E_g)$ (see the text). In (A) and (C), the curves are the results of the least-square fitting with Gaussians. In (B) and (D), the $y = 0$ information is brought from Ref. 9.

# Supporting Information.
## Quantum paramagnet near spin-state transition

*Keisuke Tomiyasu,\* Naoko Ito, Ryuji Okazaki, Yuki Takahashi, Mitsugi Onodera, Kazuaki Iwasa, Tsutomu Nojima, Takuya Aoyama, Kenya Ohgushi, Yoshihisa Ishikawa, Takashi Kamiyama, Seiko Ohira-Kawamura, Maiko Kofu, and Sumio Ishihara*

**Experimental Section**
Polycrystalline $LaCo_{1-y}Sc_yO_3$ samples were synthesized using a combination of the conventional solid-state reaction and the floating zone method. The stoichiometric mixture of $La_2O_3$ and $Co_3O_4$ was heated once at 1050 °C for 12 h and twice at 1200 °C for 24 h with thorough intermediate grinding to obtain $LaCoO_3$. The stoichiometric mixture of $La_2O_3$ and $Sc_2O_3$ was heated once at 1450 °C for 12 h and twice at 1600 °C for 24 h with thorough intermediate grinding to obtain $LaScO_3$. Next, $LaCoO_3$ and $LaScO_3$ were mixed in stoichiometric ratios, shaped into rods with a diameter of approximately 6 mm using 100 MPa hydro-pressure, sintered for 1 h at 1300 °C, and melted under $O_2$ gas flow in an image furnace. Finally, the obtained as-grown samples were annealed at 750 °C for 3 h under $O_2$ gas flow. The solid solubility limit in this method was $y \sim 0.15$. $Sr_2CoO_3Cl$ was synthesized by following the reported method.[S1]

Magnetization measurements were performed with standard superconducting quantum interference device (SQUID) magnetometers at the Center for Low Temperature Science at Tohoku University. Dilatability and magneto-lattice-expansion measurements were performed using the strain-gauge method (KFL-type from Kyowa Dengyo Co.) with the same SQUID machines. The $SiO_2$ reference was simultaneously measured.

Electrical resistivity was measured using a standard four-probe method with a current ($I$) of 10 µA, which was provided by a Keithley 6221 current source. The sample voltage was measured with a synchronized Keithley 2182A nanovoltmeter. These instruments were operated in their built-in Delta modes to cancel the unwanted thermoelectric voltage. The thermopower was measured using a steady-state technique with a temperature gradient of 0.5 K/mm and monitored with a differential thermocouple made of copper and constantan. The thermoelectric voltage of the sample was measured with the Keithley 2182A nanovoltmeter, while the voltage contribution from the wire leads was carefully subtracted.

X-ray diffraction was performed with Cu $K\alpha$ radiation diffractometers (SmartLab, Rigaku Co.). The 4 K data were recorded with a He closed-cycle refrigerator.

Neutron spectroscopy was performed on the chopper spectrometer AMATERAS (BL14) at the MLF of the J-PARC spallation neutron source (Japan).[S2] The incident energy $E_i$ was set to 4.7 meV, and the energy resolution under elastic condition was approximately 2.3% to $E_i$. The main disk chopper speed was fixed at 300 Hz. A He closed-cycle refrigerator was used. The data were obtained by the UTSUSEMI software provided by the MLF.[S3]

**Powder X-ray diffraction**

**Figure S1**A and S1B show the representative X-ray diffraction patterns measured for the $y = 0.05$ sample at $T = 298$ and 4 K, respectively. The fitting is performed using Z-Rietveld software.[S4] The data measured for all of the samples are analyzed in an $R\bar{3}c$ space group, as is typically done for the parent compound, $LaCoO_3$.[S5] The lattice volume gradually increases with $y$, as shown in the insets, ensuring that a thorough intermixing of the Co and Sc ions (solid solution) in the $LaCo_{1-y}Sc_yO_3$ samples is obtained. Furthermore, this lattice expansion is in agreement with the relations of the ionic radii, $r[Sc^{3+}] = 0.745$ Å $> r[Co^{3+(LS)}] = 0.545$ Å, $r[Co^{3+(HS)}] = 0.61$ Å, $r[Co^{3+(IS)}] = 0.56$ Å.[19,S5]

However, as written in the main text, the averaged lattice volume and the averaged $B$–O distance are below those given by Vegard's law. This suggests the contraction of the Co–O distance and the increase of the Co–O covalence, which are consistent with the relation of electronegativity, $\chi_{EN}(Sc) = 1.36 < \chi_{EN}(Co) = 1.88$. Although the short Co–O distance might seem to oppositely stabilize an LS state, the high Co–O covalence is expected to broaden the Co-3$d$-orbital bandwidth $W$ and, hence, narrow the energy gap $E_g$.[S6,S7] Thus, the Sc substitution can moderately destabilize an LS state, and is expected to generate the magnetic and highly covalent spin-state, which is characteristic in comparison to the Goodenough theory that the magnetic HS covalence is generally weaker than the LS one.[S6]

In contrast, another substitution system, $LaCo_{1-y}Rh_yO_3$, exhibits lattice expansion above Vegard's law and weak ferromagnetism comprising HS (or possibly IS) states above $y \sim 0.10$.[24,S8–S11] This behavior is consistent with the relation, $\chi_{EN}(Rh) = 2.28 > \chi_{EN}(Co) = 1.88$,[24] which may rapidly push the system up to an explicitly magnetic HS/IS state that is conventional in terms of the ionic radius. However, a relatively ionic/localized quantum-spin-state could hide in a diluted and limited range of Rh concentration.

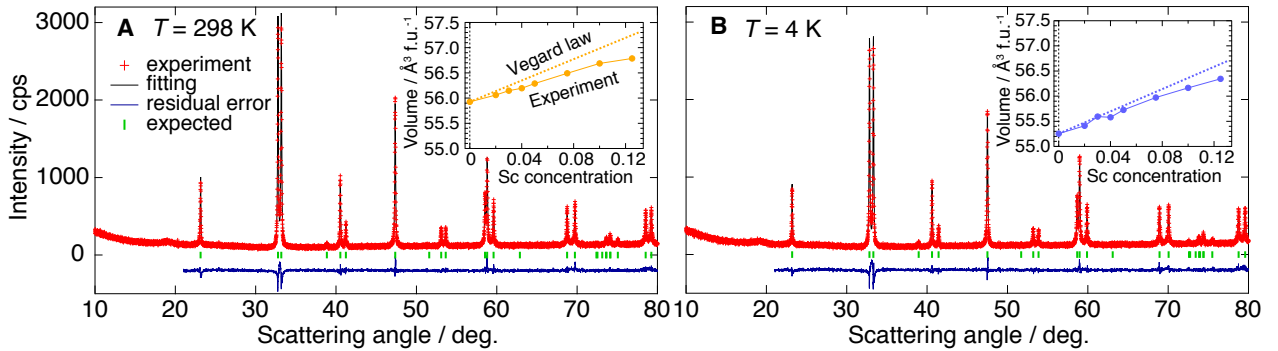

**Figure S1.** X-ray diffraction data. A) Room-temperature data. B) 4-K data. The main panels show the diffraction patterns. The insets show the $y$ dependence of the lattice volume. The dotted lines represent Vegard's law.

**Transport properties**

**Figure S2**A displays the resistivity ($\rho$) as a function of $T^{-1}$. As shown by the solid lines, the data fit well with the thermal activation formula $\rho(T) = \rho_0\exp(E_g^\rho/2k_BT)$, where $E_g^\rho$ denotes the activation energy evaluated from the resistivity data. Figure S2B depicts the thermopower ($S$) as a function of $T^{-1}$. At low temperatures, the data also exhibit the thermal activation form $S(T) = E_g^S/2qT$, where $E_g^S$ is the activation energy evaluated from the thermopower data, and $q$ is the charge of the carrier.

The parent compound LaCoO$_3$ ($y = 0$) has a value of $S$ with a positive sign. In this system, electrons and holes coexist, and the holes dominate as carriers[S12] probably because the spin-state blockade significantly suppresses the electron hopping in the Co$^{3+(LS)}$ matrix.[S13,S14] Meanwhile, $S$ becomes negative as $y$ increases most likely because the Sc substitution destabilizes the LS state of the Co$^{3+}$ and weakens the spin-state blockade effect, thereby recovering the electron hopping.

Figure 2D (the main text) shows the two activation energies obtained as functions of $y$. For the parent compound, $E_g^\rho \sim 0.37$ eV is larger than both $E_g^\rho \sim 0.21$ eV[S15] and $E_g^\rho \sim 0.23$ eV[S16] obtained from the resistivity data of single crystals probably because a mobility gap opens in the measured polycrystalline samples. Meanwhile, $E_g^S \sim 0.24$ eV is close to the values given in the literature because the thermopower is insensitive to the relaxation time; because it measures the ratio of entropy current to charge current, in which the relaxation time is canceled out. Both $E_g^S$ and $E_g^\rho$ decrease with an increasing $y$, indicating that the energy difference between valence $t_{2g}$ orbital and the relatively more conductive $e_g$ orbital decreases, resulting in the destabilization of the LS state caused by the Sc substitution.

Furthermore, $E_g^S$ and $E_g^\rho$ commonly show an interesting upturn anomaly at approximately $y = 0.04$, above which the magneto-lattice-expansion $\varepsilon(H)$ is significantly enhanced. The upturn implies a change in the electronic structure toward the insulating side at values above $y = 0.04$, which is consistent with the excitonic *insulating* model. In addition, LaCo$_{1-y}$Sc$_y$O$_3$ remains electrically not conductive, but insulating with a small energy gap (activation energy) in the present $T$ range, as shown in Figure 2C in the main text, which is also consistent with the excitonic *insulating* model.

However, the present transport measurements are limited above $T = 150$ K because the resistivity exponentially increases beyond the measurable scale in the low-$T$ range. Meanwhile, the enhancement of $\varepsilon(H)$ is obtained at $T$ values below approximately 30 K. Therefore, the observed upturn is considered a precursor of the EI state appearing in the high-$T$ range.

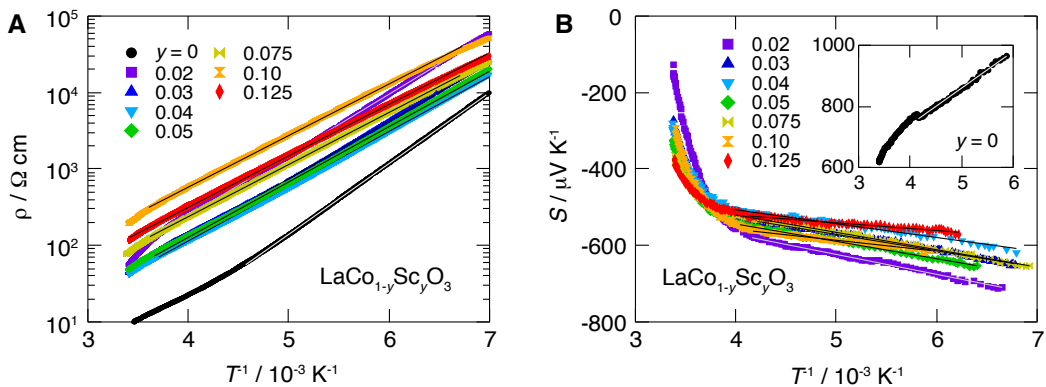

**Figure S2.** Electrical resistivity and thermopower data. A) Resistivity. B) Thermopower. The inset shows the thermopower data measured for the parent compound. In all of the panels, the symbols denote the experimental data, while the lines denote the fitting results.

**Reference data for the magneto-lattice-expansion**

**Figure S3** shows the data measured for the HS $Sr_2CoO_3Cl$ at $T = 1.8$ K, the mixed LS–HS $LaCoO_3$ at $T = 300$ K, and $LaCoO_3$ at $T = 50$ K on the verge of LS-to-HS transition. The magneto-lattice-expansion for each does not exceed $2 \times 10^{-5}$, which is a negligible value.

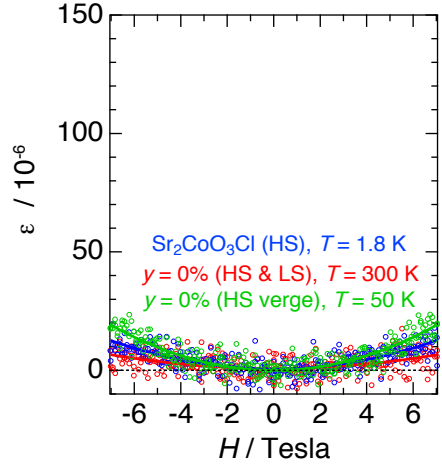

**Figure S3.** Magneto-lattice-expansion data measured for the reference compounds. The small circles denote the experimental data, while the solid curves denote the polynomial fitting.

**Relation between $\varepsilon$ and $\Delta p_{HS}$ in the conventional model of the LS and HS mixture**

The magneto-lattice-expansion at $H$ is described by:

$$\varepsilon(H) = \frac{\Delta r(H)}{r(0)}$$
$$= \frac{p_{HS}(H)r_{HS} + \{1 - p_{HS}(H)\}r_{LS} - p_{HS}(0)r_{HS} + \{1 - p_{HS}(0)\}r_{LS}}{p_{HS}(0)r_{HS} + \{1 - p_{HS}(0)\}r_{LS}}$$
$$= \frac{r_{HS} - r_{LS}}{r_{HS} + p_{HS}(0)(r_{HS} - r_{LS})}\{p_{HS}(H) - p_{HS}(0)\}$$
$$= \frac{1}{\left(1 - \frac{r_{HS}}{r_{LS}}\right)^{-1} + p_{HS}(0)} \Delta p_{HS}(H).$$

Using Shannon's ionic radii, $r_{HS} = 0.61$ Å and $r_{LS} = 0.545$ Å[19]

$$\varepsilon(H) \approx \frac{1}{8.38 + p_{HS}(0)} \Delta p_{HS}(H).$$

$0 \leq p_{HS} \leq 1$; hence, one obtains the relation equation used in the main text.

$$\varepsilon(H) \approx (0.12 \pm 0.02)\Delta p_{HS}(H).$$

**Further analysis and discussion of the quantum spin-state**

The small-$E_g$ state is accompanied by the Van Vleck effect under $H$.[25] This effect is purely quantum.[25] In a system consisting of a ground state $|0\rangle$ and an excited state $|1\rangle$ with an energy gap of $E_g$, $H$ changes the ground state from $|0\rangle$ to $|0'\rangle = |0\rangle + (\langle 1|\mu_z|0\rangle H/E_g)|1\rangle$, even when $H$ is low and $|0\rangle$ is nonmagnetic, where $\mu_z$ denotes the operator of the magnetic moment parallel to the magnetic field, which has angular momentum. However, in the case of the well-defined conventional states described by

$$\begin{cases} |0\rangle = |\text{LS}\rangle \\ |1\rangle = |\text{HS}\rangle \end{cases}$$

corresponding to the parent LaCoO$_3$, the coefficient $\langle 1|\mu_z|0\rangle = \langle \text{HS}|\mu_z|\text{LS}\rangle = 0$ because the LS states are zero in all of the angular momenta of the spin $S$, orbital $L$, and total $J$. In fact, we observed that the LS-reference of LaCoO$_3$ at $T = 1.8$ K demonstrates virtually no response to $H$ for both $M$ and $\varepsilon$. In contrast, in the case of the LS- and HS-based quantum states described by

$$\begin{cases} |0\rangle = |\text{LS-Q}\rangle = |\text{LS}\rangle + \alpha|\text{HS}\rangle \ (|\alpha| \ll 1) \\ |1\rangle = |\text{HS-Q}\rangle = |\text{HS}\rangle + \beta|\text{LS}\rangle \ (|\beta| \ll 1) \end{cases},$$

the coefficient $\langle 1|\mu_z|0\rangle = \alpha\langle \text{HS}|\mu_z|\text{HS}\rangle = \alpha\mu_{\text{HS}} \neq 0$. This finite quantum Van Vleck effect induced by weak $H$ is distinctive of the quantum-superposition states in contrast to the conventional LS and HS states.

We attempt to estimate the values of $E_g$ and $|\alpha|$. First, with regard to the $E_g$ value, we explain the analysis of the experimental $T$ dependence of $\varepsilon$(7T), as shown in the inset of Figure 3B. The magneto-lattice-expansion at $H = H_0$ is described by

$$\begin{cases} \varepsilon(T, H_0) = p_0(T)\varepsilon_0(H_0) + p_1(T)\varepsilon_1(H_0) \propto p_0(T) + cp_1(H_0) \\ p_0(T) = \dfrac{1}{\left\{1 + \exp\left(-\dfrac{E_g}{k_B T}\right)\right\}}, \\ p_1(T) = 1 - p_0(T) \end{cases}$$

where $p_0(T)$ and $p_1(T)$ denote the statistical populations of the ground and excited states, respectively; $\varepsilon_0(H_0)$ and $\varepsilon_1(H_0)$ represent their positive and negative magneto-lattice-expansion, respectively; and $c$ represents the ratio $\varepsilon_1(H_0)/\varepsilon_0(H_0)$. Thus, although the calculations for $\varepsilon_0(H_0)$ and $\varepsilon_1(H_0)$ require detailed descriptions for both the ionic-radius operators and the wave functions of the states, the curve profile of the $T$ dependence of $\varepsilon(H_0)$ can be analyzed without them to provide the $E_g$ value. The best-fitting value is $E_g = 2.9$ meV.

This value is consistent with the neutron result of $\langle E_g \rangle = 2.0$ meV. However, the neutron study also indicates that the $E_g$ values are slightly distributed to the negative side. These minor Co sites exhibit the ground HS-based quantum states, which will be accompanied by the opposite trend to the ground LS-based quantum states. The LS characteristics recover with an increasing $H$, and the negative magneto-lattice-expansion will be expected. Thus, it is considered that the partly canceled positive magneto-lattice-expansion is observed in total.

Next, with regard to the $|\alpha|$ value, the Van Vleck paramagnetic susceptibility ($\chi_{VV}$) is derived as follows:[25] the HS state, $|\text{HS}\rangle$, is described by an effective total angular momentum $J_{\text{eff}} = 1$ and, hence, consists of the three states $J_{\text{eff},z} = 1, 0,$ and $-1$. Therefore, we more accurately describe the LS-based EI states, $|0\rangle$ without $H$ and $|0'\rangle$ under $H$, as:

$$|0\rangle = |LS\rangle + \alpha(|HS,+1\rangle + |HS,0\rangle + |HS,-1\rangle)$$

and

$$|0'\rangle = |0\rangle + (\alpha\mu_{HS}H/E_g)\{1\cdot|HS,+1\rangle + 0\cdot|HS,0\rangle + (-1)\cdot|HS,-1\rangle\}$$
$$= |0\rangle + (\alpha\mu_{HS}H/E_g)(|HS,+1\rangle - |HS,-1\rangle),$$

where $\mu_{HS} = gJ_{eff}\mu_B$. The change in magnetization induced by $H$ is calculated by:

$$\Delta M(H) = \langle 0'|\hat{\mu}_z|0'\rangle - \langle 0|\hat{\mu}_z|0\rangle$$
$$= 4|\alpha|^2\mu_{HS}^2 H/E_g + O(H^2),$$

where $\langle HS,\pm1|\mu_z|HS,\pm1\rangle = \pm\mu_{HS}$ and $\langle HS,0|\mu_z|HS,0\rangle = 0$ are used. Thus, one obtains the Van Vleck paramagnetic susceptibility at the lowest possible temperature as:

$$\chi_{VV,cal} = \Delta M(H)/H = 4|\alpha|^2\mu_{HS}^2/E_g.$$

Meanwhile, an experimental value of the linear magnetization slope, $\chi_{VV,exp} = 2.8 \times 10^{-3}/\mu_B$ f.u.$^{-1}$ T$^{-1}$, is approximately observed at $T = 1.8$ K for the $y = 0.04$ sample (Figure 2E). For $\mu_{HS} = gJ_{eff}\mu_B$, we used HS values of $g = 3.4$ and $J_{eff} = 1$, which were experimentally determined for LaCoO$_3$.[9,10,22] Thus, $|\alpha| = 0.06$ is roughly estimated by using the equality $\chi_{VV,cal} = \chi_{VV,exp}$ and the aforementioned $E_g = 2.9$ meV. This supports the claim of $|\alpha| \ll 1$.

In the Van Vleck model, the lattice-expansion induced by $H$ is calculated by:

$$\varepsilon(H) = (\langle 0'|r|0'\rangle - \langle 0|r|0\rangle)/\langle 0|r|0\rangle,$$

where $r$ denotes the operator of ionic radius. The $H^1$ term is canceled out, and the $H^2$ term is the lowest order, which is consistent with the experimental data (Figure 3). However, the perturbative analytical expression becomes very complicated, and the value intricately depends on the full set of complex coefficients ($\alpha_1, \alpha_0, \alpha_{-1}; \beta_1, \beta_0, \beta_{-1}$) describing the quantum states,

$$|0\rangle = |LS\rangle + \alpha_1|HS,+1\rangle + \alpha_0|HS,0\rangle + \alpha_{-1}|HS,-1\rangle.$$
$$\begin{cases} |1,+1\rangle = |HS,+1\rangle + \beta_1|LS\rangle \\ |1,0\rangle = |HS,0\rangle + \beta_0|LS\rangle \\ |1,-1\rangle = |HS,-1\rangle + \beta_{-1}|LS\rangle \end{cases}.$$

Thus, too many free parameters can be used for a unique determination at this stage. However, in turn, this parametric flexibility allows reproducing a quite arbitrary value of $\varepsilon(H)$. In fact, by numerical trials and errors using exact diagonalization, we found parameter sets that satisfy both the observed values for $y = 0.04$, $\chi_{VV,exp} = 2.8 \times 10^{-3}/\mu_B$ f.u.$^{-1}$ T$^{-1}$ and $\varepsilon(7T) = 9 \times 10^{-5}$.

Theoretically, if the phases of coefficients form site-coherence, the superconductivity-like state will emerge, in which the number of HS cannot be defined (bi-EI state). If not so, the pseudo-gap-like incoherent precursors will occur. These are intriguing issues of the future.

**Matrix elements used for the neutron data analysis**
The matrix elements among the conventional LS and HS states are described by

$$|\langle \mathrm{LS}|M_\perp|\mathrm{LS}\rangle|^2 = 0,$$
$$|\langle \mathrm{HS}|M_\perp|\mathrm{LS}\rangle|^2 = 0,$$
$$|\langle \mathrm{LS}|M_\perp|\mathrm{HS}\rangle|^2 = 0,$$
$$|\langle \mathrm{HS}|M_\perp|\mathrm{HS}\rangle|^2 = m_{\mathrm{HS}}{}^2.$$

The second relation indicates that the transition from LS to HS cannot be observed because the LS states are zero in all of the angular momenta of spin *S*, orbital *L*, and total *J*. This explains why no signal is observed at the lowest temperatures in the *y* = 0 sample. Meanwhile, the triply degenerate HS states are split by small rhombohedral crystalline field (Figure 5D). The internal HS transition can be observed at the 0.6 meV excitations when the HS states are thermally activated.[9]

Those among the LS- and HS-based quantum states are described by

$$|\langle \mathrm{LS\text{-}Q}|M_\perp|\mathrm{LS\text{-}Q}\rangle|^2 = |(\langle \mathrm{LS}| + \alpha\langle \mathrm{HS}|)M_\perp(|\mathrm{LS}\rangle + \alpha|\mathrm{HS}\rangle)|^2 = |\alpha|^4|\langle \mathrm{HS}|M_\perp|\mathrm{HS}\rangle|^2 \sim 0,$$
$$|\langle \mathrm{HS\text{-}Q}|M_\perp|\mathrm{LS\text{-}Q}\rangle|^2 = |(\langle \mathrm{HS}| + \beta\langle \mathrm{LS}|)M_\perp(|\mathrm{LS}\rangle + \alpha|\mathrm{HS}\rangle)|^2 = |\alpha|^2|\langle \mathrm{HS}|M_\perp|\mathrm{HS}\rangle|^2 \sim 0,$$
$$|\langle \mathrm{LS\text{-}Q}|M_\perp|\mathrm{HS\text{-}Q}\rangle|^2 = |(\langle \mathrm{LS}| + \alpha\langle \mathrm{HS}|)M_\perp(|\mathrm{HS}\rangle + \beta|\mathrm{LS}\rangle)|^2 = |\alpha|^2|\langle \mathrm{HS}|M_\perp|\mathrm{HS}\rangle|^2 \sim 0,$$
$$|\langle \mathrm{HS\text{-}Q}|M_\perp|\mathrm{HS\text{-}Q}\rangle|^2 = |(\langle \mathrm{HS}| + \beta\langle \mathrm{LS}|)M_\perp(|\mathrm{HS}\rangle + \beta|\mathrm{LS}\rangle)|^2 = |\langle \mathrm{HS}|M_\perp|\mathrm{HS}\rangle|^2 = m_{\mathrm{HS}}{}^2.$$

These are equal to the aforementioned ones when $|\alpha| \ll 1$.